\documentclass[journal,onecolumn,12pt]{IEEEtran}

\usepackage{amssymb,amsmath,amsfonts,bm,epsfig,graphicx,theorem,latexsym}
\usepackage{rotating,setspace,latexsym,epsf,color,subfigure}
\usepackage{cite,authblk}

\newtheorem{lemma}{Lemma}

\newenvironment{Proof}[1]{\medskip\par\noindent
{\bf Proof:\,}\,#1}{{\mbox{\,$\blacksquare$}\par}}

\allowdisplaybreaks

\begin{document}
\IEEEoverridecommandlockouts

\title{Energy Harvesting Transmitters that Heat Up: Throughput Maximization under Temperature Constraints}

\author{Omur Ozel \qquad Sennur Ulukus \qquad Pulkit Grover\thanks{Omur Ozel was with the Department of Electrical and Computer Engineering, University of Maryland, College Park, MD 20742. He is now with the Department of Electrical Engineering and Computer Sciences, University of California, Berkeley, CA 94720. Sennur Ulukus is with the Department of Electrical and Computer Engineering, University of Maryland, College Park, MD 20742. Pulkit Grover is with the Department of Electrical and Computer Engineering, Carnegie Mellon University, Pittsburgh, PA 15213. Emails: ozel@berkeley.edu, ulukus@umd.edu, pgrover@andrew.cmu.edu. This work was supported by NSF Grants CNS 13-14733, CCF 14-22111, CCF 14-22129, CCF 13-50314, ECCS 13-43324, and presented in part at the IEEE International Symposium on Information Theory, Hong Kong, June 2015.}}

\maketitle
\vspace{-1.2cm}
\begin{abstract}
Motivated by damage due to heating in sensor operation, we consider the throughput optimal offline data scheduling problem in an energy harvesting transmitter such that the resulting temperature increase remains below a critical level. We model the temperature dynamics of the transmitter as a linear system and determine the optimal transmit power policy under such temperature constraints as well as energy harvesting constraints over an AWGN channel. We first derive the structural properties of the solution for the general case with multiple energy arrivals. We show that the optimal power policy is piecewise monotone decreasing with possible jumps at the energy harvesting instants. We derive analytical expressions for the optimal solution in the single energy arrival case. We show that, in the single energy arrival case, the optimal power is monotone decreasing, the resulting temperature is monotone increasing, and both remain constant after the temperature hits the critical level. We then generalize the solution for the multiple energy arrival case.

\end{abstract}

\section{Introduction}

In many wireless sensor applications, temperature increase caused by sensor operation has to be carefully managed. For example, wireless sensors implanted in the human body have to be designed such that the temperature due to their operation does not cause any threat for the metabolism. A line of medical research started by Pennes in 1948 \cite{pennes48} explores the temperature dynamics due to electromagnetic radiation in conjunction with heat losses to the environment and dissipation of heat in the tissue. In the context of sensors that communicate data, temperature sensitivity varies depending on the type of tissue. For a given specific tissue, it is recommended that the temperature does not exceed a critical level, in order to prevent damage to the tissue. This necessitates careful scheduling of data transmission \cite{scheduling_bio2005}. This problem arises in various types of body area sensor networks, see e.g., \cite{bsn2,survey, thermal_survey} and references therein. Finally, temperature increase in a sensor is a threat for the proper operation of the hardware itself \cite{thermal,thermal_conf,thermal2,ankur13}. In this context, the electric power that feeds the amplifier circuitry has to be carefully scheduled so as to avoid permanent damage in the circuit.

In order to obtain design principles with regard to temperature sensitivity of such systems, determining transmission schemes under a safe temperature threshold $T_c$ is a useful objective. In this paper, we consider data transmission with energy harvesting sensors under such temperature constraints. Data transmission with energy harvesting transmitters has been the topic of recent research \cite{tcom-submit,kaya_subm,ozel11,zhang_tsp, finite, jing12jcn, kaya_jcn, mitran_isit}. In particular, throughput maximization under offline and online knowledge of the energy arrivals is considered in these references for single-user and multi-user energy harvesting communication systems. In \cite{Gunduz-leakage,kaya_storage,orhan_twc, zhang_process,ozel_tsp}, this problem is investigated under imperfections such as battery energy leakage, charge/discharge inefficiency, and presence of processing costs.

In the current paper, we aim to bridge physical heat dissipation with data transmission in energy harvesting communication systems. When the sole purpose is to maximize the throughput, the transmitter may generate excessive heat while utilizing the energy resource. In a temperature sensitive application, the heat accumulation caused by the transmission power policy has to be explicitly taken into account. In such a case, heat generated in the transmitter circuitry causes a form of ``information-friction" \cite{pulkit_friction}. We study the effect of this ``friction" in a deadline constrained communication of an energy harvesting transmitter over an AWGN channel. For simplicity, we use transmit power as a proxy for hardware power. That is, we assume that the energy dissipated by the power amplifier dominates other energy sinks in the circuitry. More work is needed to understand full implications of communication circuitry's energy in this context. Our formulation also relates to \cite{channels_heat} in that the cumulative effect of heat generated in the hardware affects the communication performance.

We determine the throughput optimal offline power scheduling policy under energy harvesting and temperature constraints. Our thermal model is based on a view of the transmitter's circuitry as a linear heat system where transmit power is an input as in \cite{pennes48}, \cite{thermal_conf,ankur13}, \cite{channels_heat}. We impose that the temperature does not exceed a critical level $T_c$. Consequently, we obtain a convex optimization problem. We solve this problem using a Lagrangian framework and KKT optimality conditions. We first derive the structural properties of the solution for the general case of multiple energy arrivals. Then, we obtain closed form solutions under a single energy arrival. For the general case, we observe that the optimal power policy may make jumps at the energy arrival instants, generalizing the optimal policies in \cite{tcom-submit,kaya_subm}. Between energy harvests, the optimal power is monotonically decreasing. We establish for the case of a single energy arrival that the optimal power policy monotonically decreases, corresponding temperature monotonically increases, and both remain constant when the critical temperature is reached. Then, we consider the case of multiple energy arrivals. We observe that the properties of the solution for the single energy arrival case are guaranteed to hold only in the last epoch of the multiple energy arrival case. In the remaining epochs, the temperature may not be monotone and the transmitter may need to cool down to create a temperature margin for the future, if the energy harvested in the future is large. We illustrate possible cases and obtain insights regarding the optimal temperature pattern in the multiple energy arrival case.

\section{The Model}
\label{sect_model}

\begin{figure}[t]
\centerline{\includegraphics[width=0.6\linewidth]{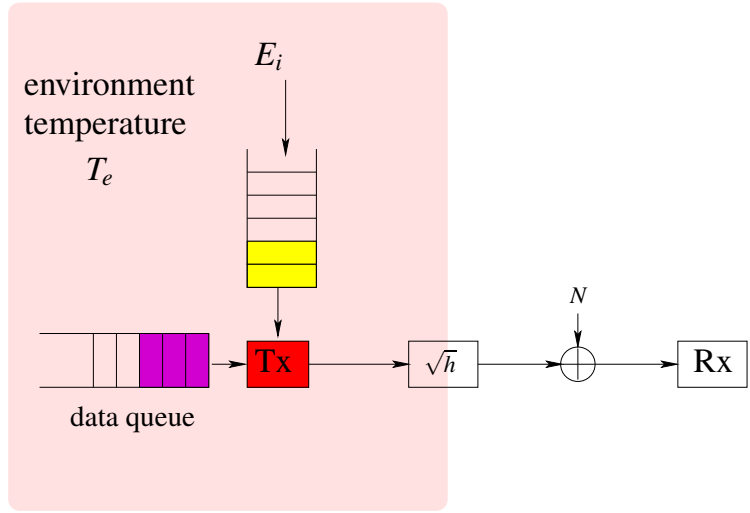}}
\caption{The model representing an energy harvesting wireless node placed in an environment that has constant temperature $T_e$.}
\label{model}
\end{figure}

We consider an energy harvesting transmitter node placed in an environment as depicted in Fig.~\ref{model}. The node harvests energy to run its circuitry and wirelessly send data to a receiver.

\subsection{Channel Model}
The received signal $Y$, the input $X$, fading level $h$ and noise $Z$ are related as
\begin{align}
Y=\sqrt{h}X + Z
\end{align}
where $Z$ is additive white Gaussian noise with zero-mean and unit-variance. In this paper, the channel is non-fading, i.e., $h=1$. We use a continuous time model: A scheduling interval has a short duration with respect to the duration of transmission and we approximate it as $[t,t+dt]$ where $dt$ denotes infinitesimal time. In $[t,t+dt]$, the transmitter decides a feasible transmit power level $P(t)$ and $\frac{1}{2}\log\left(1+P(t)\right)dt$ bits are sent to the receiver, where the base of $\log$ is $2$. To be precise, the underlying physical signaling is in discrete time and the scalings in SNR and rate due to bandwidth and the base of the logarithm are inconsequential for the analysis.

\subsection{Energy Harvesting Model}

\begin{figure}[t]
\centerline{\includegraphics[width=0.75\linewidth]{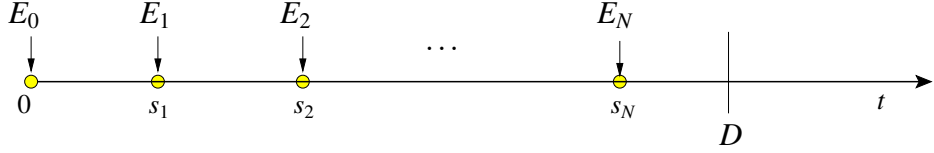}}
\caption{Energy $E_i$ becomes available for data transmission at time $s_i$. $D$ is the deadline. }
\label{model2}
\end{figure}

As shown in Fig.~\ref{model2}, the initial energy available in the battery at time zero is $E_0$. Energy arrivals occur at times $\{s_1,s_2,\ldots\}$ in amounts $\{E_1,E_2,\ldots\}$ with $s_0=0$. We call the time interval between two consecutive energy arrivals an {\it epoch}. $D$ is the deadline. $E_i$ and $s_i$ are known offline and are not affected by the heat due to transmission. Let $h(t)=\max\{k: s_k < t\}$ and $N$ be the number of energy arrivals in the interval $[0,D)$ and by convention we let $s_{N+1}=D$. Power scheduling policy $P(t)$ is subject to energy causality constraints as:
\begin{align}\label{causality}
\int_{0}^{t} P(\tau) d\tau \leq \sum_{i=0}^{h(t)} E_i, \qquad \forall t \in [0,D]
\end{align}

\subsection{Thermal Model}

In our thermal model, we use the transmit power as a measure of heat dissipated to the environment. In particular, we model the temperature dynamics of the system as follows:
\begin{align}
\label{thermal}
\frac{d }{dt}T(t) = a P(t) - b (T(t) - T_e) + c
\end{align}
where $P(t)$ is the transmit power policy and $T(t)$ is the temperature at time $t$. $T_e$ is the constant temperature of the environment that is not affected by the heating effect due to the transmit power level $P(t)$. $a$ and $b$ are non-negative constants. $c$ represents the cumulative effect of additional heat sources and sinks and it can take both positive and negative values. In the following, we consider the case of no extra heat source or sink, i.e., $c=0$.

Our thermal model in (\ref{thermal}) is intimately related to the thermal model in \cite{thermal_conf, ankur13} where hardware heating is modeled as a first order $RC$ heat circuit. In particular, thermal dynamics of a power controlled transmitter due to its amplifier power consumption (see e.g., \cite{cui_goldsmith05}) could be modeled as in (\ref{thermal}). We also refer the reader to \cite{channels_heat} for a related heating model. Our thermal model is also related to the well-known Pennes bioheat equation \cite{pennes48}. We assume, for simplicity, that the spatial variation in temperature is not significant and leave the general case of spatial temperature variations as future work.

From (\ref{thermal}), the solution of $T(t)$ for any given $P(t)$ with the initial condition $T(t')$ at time $t'$ is:
\begin{align}\label{four}
\hspace{-0.12in}T(t) = e^{-b(t-{t'})} \left(\int_{t'}^t e^{b(\tau-t')} \left(a P(\tau) + b T_e\right) d\tau + T(t') \right)
\end{align}
By inserting $t'=0$ in (\ref{four}), we get (c.f. \cite[Eq. (3)]{channels_heat}):
\begin{align}\label{ths}
T(t) = e^{-bt} \left(\int_0^{t} e^{b \tau} \left(aP(\tau) + bT_e\right) d \tau + T(0) \right)
\end{align}

The temperature should remain below a critical temperature $T_c$, i.e., $T(t) \leq T_c$, where we assume that $T_c > T_e$. Let us define $T_{\delta} \triangleq T_c - T_e$, which is the largest allowed temperature deviation from the environment temperature. Typically, initial temperature is $T_e$, i.e., initially the temperature is stabilized at the constant environment temperature $T_e$. From (\ref{ths}), using $T(t)\leq T_c$ and $T(0)=T_e$, we get the following equivalent condition for the temperature constraint:
\begin{align} \label{gg}
\int_0^{t} a e^{b \tau} P(\tau) d \tau \leq T_{\delta}e^{bt}, \quad \forall t \in [0,D]
\end{align}
Note that the temperature constraints in (\ref{gg}) and the energy causality constraints in (\ref{causality}) do not interact. Due to the heat generation dynamics governed by (\ref{thermal}), we observe in (\ref{gg}) that the cost of power increases exponentially in time (i.e., the multiplier in front of $P(\tau)$ is exponential in $\tau$) while the heat budget also increases exponentially in time (i.e., the upper bound on the right hand side of (\ref{gg}) is exponential in $t$).

\section{Problem Formulation}
\label{sect_for}

Offline throughput maximization problem over the interval $[0,D]$ under energy causality and temperature constraints with initial temperature $T(0)=T_e$ is:
\begin{align}\nonumber
\max_{P(t), \ t \in [0,D]} \quad &\int_0^D \frac{1}{2}\log\left(1 + P(\tau)\right)d\tau \\ \nonumber
\mbox{s.t.} \quad &\int_0^{t} a e^{b \tau} P(\tau) d \tau \leq T_{\delta}e^{bt}, \quad  \forall t \\
\qquad &\int_{0}^{t} P(\tau) d\tau \leq \sum_{i=0}^{h(t)} E_i, \quad \forall t \label{opt_prob1}
\end{align}
where the space of actions is the set of measurable functions $P(t)$ defined over the interval $[0,D]$. Note that (\ref{opt_prob1}) is a convex functional optimization problem.

The Lagrangian for (\ref{opt_prob1}) is:
\begin{align}\nonumber
\mathcal{L}=&\int_0^D \frac{1}{2}\log\left(1+P(t)\right)dt -\int_0^D \lambda(t) \left(\int_0^{t} a e^{b \tau} P(\tau) d \tau - T_{\delta}e^{bt}\right) dt \\ & -\int_0^D \beta(t)\left( \int_{0}^{t} P(\tau) d\tau - \sum_{i=0}^{h(t)} E_i \right) dt \label{lgrng}
\end{align}
Taking the derivative of the Lagrangian with respect to $P(t)$ and equating to zero:
\begin{align}\label{rf}
\frac{1}{1+P(t)} - e^{b t} \int_t^D \lambda(\tau) d\tau - \int_{t}^D \beta(\tau)d\tau  = 0
\end{align}
which gives
\begin{align}\label{nn}
P(t) = \left[ \frac{1}{\int_t^D\beta(\tau)d\tau + e^{bt}\int_t^D \lambda(\tau)d\tau} - 1 \right]^+
\end{align}
In addition, the complementary slackness conditions are:
\begin{align}\label{slck1}
\lambda(t)\left( \int_0^{t} a e^{b \tau} P(\tau) d \tau - T_{\delta}e^{bt} \right) &= 0, \quad \forall t \\ \label{slck2}
\beta(t)\left( \int_{0}^{t} P(\tau) d\tau - \sum_{i=0}^{h(t)} E_i \right) &= 0, \quad \forall t
\end{align}
In (\ref{rf}) and (\ref{slck1})-(\ref{slck2}), $\lambda(t) \geq 0$ and $\beta(t) \geq 0$ are distributions that are allowed to have impulses and their total measure over $[0,D]$ interval are not both zero, i.e., $ \int_0^D \lambda(\tau) d\tau >0$ or $ \int_0^D \beta(\tau) d\tau > 0$, in order to prohibit $P(t)$ from being unbounded. We note that (\ref{rf}) and (\ref{slck1})-(\ref{slck2}) are necessary and sufficient conditions since the problem is convex. The solution is unique almost everywhere as the objective function is strictly concave.

We note that the problem in (\ref{opt_prob1}) could be solved by using calculus of variations. See \cite{thermal_conf} for application of calculus of variations for a similar problem to (\ref{opt_prob1}). As another alternative, we note that (\ref{opt_prob1}) could equivalently be solved by using a Hamiltonian approach from optimal control theory. In particular, we can cast the problem in (\ref{opt_prob1}) as an optimal control problem with pure state constraints \cite{opt_survey}. In this case, the state of the system is the tuple $[T(t) \ B(t)]$ where $B(t)=\int_0^t P(\tau)d\tau$ is the total energy expenditure by the time $t$. The input is $P(t)$ for $0 \leq t \leq D$. This problem is in the following form:
\begin{align}\nonumber
\max_{P(t), \ t \in [0,D]}  \quad &\int_0^D \frac{1}{2}\log\left(1 + P(\tau)\right)d\tau \\ \nonumber
\mbox{s.t.} \quad &\frac{d}{dt}T(t) = f_1(T,B,P), \ \frac{d}{dt}B(t)= f_2(T,B,P) \\
\qquad & g_1(T,B,t) \leq 0, \ g_2(T,B,t)  \leq 0 \label{opt_prob2}
\end{align}
where $f_1(T,B,P)=aP-b(T-T_e)$ and $f_2(T,B,P)=P$ while $g_1(T,B,t)=T-T_c$ and $g_2(T,B,t)=B - \sum_{i=0}^{h(t)} E_i$. Note that $g_1$ and $g_2$ do not depend on the input $P$. With these selections, optimization problem (\ref{opt_prob2}) is in the same form as that stated in \cite[Eqs. (2.1)-(2.6)]{opt_survey}. In this case, Hamiltonian is
\begin{align}
\mathcal{H}(T,B,P,\lambda_1,\lambda_2,t) = \frac{1}{2}\log\left(1 + P\right) - \lambda_1(t)f_1(T,B,P) - \lambda_2(t)f_2(T,B,P)
\end{align}
and the corresponding Lagrangian is
\begin{align}
\mathcal{L}_H(T,B,P,\lambda_1,\lambda_2,t)=\mathcal{H}(T,B,P,\lambda_1,\lambda_2,t)-\nu_1(t)g_1(T,B,t)-\nu_2(t)g_2(T,B,t)
\end{align}
where $\lambda_1(t)$ and $\lambda_2(t)$ are the co-state trajectories; $\nu_1(t)$ and $\nu_2(t)$ are multiplier functions. We note that Pontryagin's maximum principle is necessary and sufficient in this case since (\ref{opt_prob2}) is a concave maximization problem. One can derive the equivalence of necessary and sufficient conditions for this optimal control problem to those in (\ref{rf}) and (\ref{slck1})-(\ref{slck2}).

In the following, we proceed with the Lagrangian formulation in (\ref{lgrng}) and the corresponding optimality conditions in (\ref{rf}) and (\ref{slck1})-(\ref{slck2}).

\section{General Properties of an Optimal Policy }
\label{sect_prop}

In this section, we obtain the structural properties of the optimal power scheduling policy using the optimality conditions. In the following lemmas, $P(t)$ refers to the optimal policy and $T(t)$ is the resulting temperature unless otherwise stated.

We first note that the temperature level never drops below $T_e$. In particular, if the initial temperature is between $T_e$ and $T_c$, the temperature at all times will remain between $T_e$ and $T_c$.
\begin{lemma}\label{temp}
$T_e \leq T(t) \leq T_c$ whenever the initial temperature is $T_e \leq T(0) \leq T_c$.
\end{lemma}
\begin{Proof}
From (\ref{thermal}), since $P(t)\geq 0$ we have $\frac{d}{dt}T(t) \geq 0$ whenever $T(t)=T_e$. The constraint $T(t) \leq T_c$ is satisfied by any feasible policy in (\ref{opt_prob1}).
\end{Proof}

The following lemma states that if the temperature $T(t)$ is constant, then the power $P(t)$ is constant also (while it is not true the other way around, see Lemma~\ref{piecewise}), and that if the temperature hits the maximum allowed level $T_c$, then the power must be below a threshold.
\begin{lemma}\label{constant}
Whenever $T(t)$ is constant over an interval $I \ \subseteq \ [0,D]$, $P(t)$ is also constant over that interval. If the temperature hits the level $T_c$ at $t=t_h$, then $P(t_h+\epsilon) \leq \frac{T_{\delta}b}{a}$ for all sufficiently small $\epsilon>0$.
\end{lemma}
\begin{Proof}
If $T(t)$ is constant in $I$, $\frac{d}{dt}T(t)=0$ and from (\ref{thermal}), $P(t)$ is also constant in the same interval. If $T(t_h)=T_c$ for some $t_h \in [0,D)$, then $\frac{d}{dt}T(t_h+\epsilon) \leq 0$ and from (\ref{thermal}), $P(t_h + \epsilon) \leq \frac{T_{\delta}b}{a}$.
\end{Proof}

The following lemma shows that if the power $P(t)$ is a monotone increasing function, then so is the temperature $T(t)$. We first prove this result for piecewise constant functions and then generalize it to arbitrary functions. We note that a particular instance of a monotone increasing piecewise constant power is observed in the solution of the throughput maximization problem without temperature constraints \cite{tcom-submit}.
\begin{lemma}
\label{piecewise}
If $P(t)$ is a monotone increasing piecewise constant function, then $T(t)$ is monotone increasing. More generally, if $P(t)$ is a monotone increasing function, so is $T(t)$.
\end{lemma}
\begin{Proof}
We first prove the first statement of the lemma which is concerned with piecewise constant functions. Let us start with the case of a single constant power value for the entire duration of communication, i.e., $P(t)=p$ for $t \in [0,D]$. From (\ref{ths}), we have:
\begin{align}
T(t) &= e^{-bt} \left(\int_{0}^t e^{b\tau} \left(a p + b T_e\right) d\tau + T(0) \right)\\&= e^{-bt}\left( \frac{\left(a p + b T_e\right) }{b}\left(e^{bt}-1\right)+T(0)\right) \\ &= T_e + \frac{a}{b}p + \left( T(0) - T_e - \frac{a}{b}p\right)e^{-bt}\label{fin}
\end{align}
For $T(0)=T_e$, (\ref{fin}) is a monotone increasing function of $t$. In particular, $T(t) \leq T_e + \frac{a}{b}p$. Now, let us consider the case of $M$ constant power levels for the duration of communication, i.e., $P(t)=p_i$ over the interval $[I_{i-1},I_{i})$ where $p_{i} < p_{i+1}$ for all $i$ and $0 = I_0 < I_1 < \ldots < I_M = D $ where $M >1$ is the number of intervals. In this case, we have for $t \in [I_{i-1},I_i)$:
\begin{align}\label{fin2}
T(t)=T_e + \frac{a}{b}p_i + \left(T(I_{i-1}) - T_e - \frac{a}{b}p_i\right)e^{-b(t-I_{i-1})}
\end{align}
where $T(I_{i-1}) \leq T_e + \frac{a}{b}p_{i-1}$. Hence, the coefficient of $e^{-b(t-I_{i-1})}$ in (\ref{fin2}) has a negative sign as $T(I_{i-1}) - T_e - \frac{a}{b}p_i \leq \frac{a}{b}\left(p_{i-1}-p_i\right) < 0$. This proves that $T(t)$ is monotone increasing.

To generalize this result for any monotone increasing function $P(t)$, we obtain any monotone increasing simple approximation \cite{fitzpatrick10} of $P(t)$, denoted as $P_n(t)$, such that $P_1(t) \leq P_2(t) \leq \ldots \leq P_n(t)$ for all $t \in [0,D]$ and $P_n(t) \rightarrow P(t)$ pointwise. For example, one can select $P_n(t)=P(I_{n(i-1)})$ for $t \in [I_{n(i-1)}, I_{ni})$ and $I_{ni}=\frac{D}{2^n}\left(i-1\right)$ for $i=1, \ldots, 2^{n}$. Let us call the resulting temperature $T_n(t)$. Hence, $e^{bt}P_1(t) \leq e^{bt}P_2(t) \leq \ldots \leq e^{bt} P_n(t)$ for all $t \in [0,D]$ and $e^{bt} P_n(t) \rightarrow e^{bt} P(t)$ pointwise. By monotone convergence theorem \cite{fitzpatrick10}, we have
\begin{align}
\int_0^t e^{b\tau}P_n(\tau)d\tau \rightarrow \int_0^t e^{b\tau}P(\tau)d\tau, \quad \forall t \in [0,D]
\end{align}
Accordingly, $T_n(t)\rightarrow T(t)$ pointwise and we have
\begin{align}
\frac{d}{dt} T_n(t) = aP_n(t) - b\left(T_n(t) - T_e\right) \rightarrow \frac{d}{dt} T(t) = aP(t) - b\left(T(t) - T_e\right), \ \forall t \in [0,D]
\end{align}
Since $P_n(t)$ is a monotone increasing piecewise constant function, from the first part of the proof, $T_n(t)$ is monotone increasing, i.e., $\frac{d}{dt} T_n(t) = aP_n(t) - b\left(T_n(t) - T_e\right) \geq 0$. Since $\frac{d}{dt} T_n(t) \rightarrow \frac{d}{dt} T(t)$ pointwise, this implies $\frac{d}{dt} T(t) \geq 0$, i.e., $T(t)$ is monotone increasing as well.
\end{Proof}

The next lemma shows that if the temperature remains constant over an interval, then that level could only be $T_e$ or $T_c$, i.e., any other temperature cannot be a stable temperature.
\begin{lemma}
\label{const_temp}
If $T(t)$ is constant over an interval $I \ \subseteq \ [0,D]$, then that constant level could only be $T_e$ or $T_c$.
\end{lemma}
\begin{Proof}
Assume $T(t)$ is constant over $I$. Without loss of generality, assume that there is no energy arrival in the interval $I$, and otherwise let $I$ be the portion of the interval without any energy arrivals. By Lemma~\ref{constant}, $P(t)$ is constant over $I$. If $P(t)=0$ over $I$, then $T(t)=T_e$ from (\ref{thermal}). If $P(t)\neq 0$, we have from (\ref{nn})
\begin{align}\label{nn2}
P(t) = \frac{1}{\int_t^D\beta(\tau)d\tau + e^{bt}\int_t^D \lambda(\tau)d\tau} - 1
\end{align}
where $\beta(t)=0$ over the interval $I$ by (\ref{slck2}) since $\beta(t)>0$ implies energy constraint is tight and $P(t)=0$. Therefore, $\int_t^D \beta(\tau)d\tau=B$ is constant over $I$. If $T(t)<T_c$, then by (\ref{slck1}), $\lambda(t)=0$ over $I$ and hence $\int_t^D \lambda(\tau)d\tau = C$ is constant over $I$. However, this makes (\ref{nn2}) a time varying function of $t$ because of the $e^{bt}$ term in the denominator, and this contradicts the fact that $P(t)$ is constant. Finally, if $C=0$, this means that the temperature constraint is never tight. In this case, the piecewise constant power policy in \cite{tcom-submit} is optimum, and the temperature is monotonically increasing from Lemma~\ref{piecewise}, and therefore, cannot be a constant over an interval.
\end{Proof}

The following lemma states that at the end of the communication session either the harvested energy is exhausted or the critical temperature is reached.
\begin{lemma}\label{lm12}
At $t=D$, either the temperature constraint or the energy causality constraint or both are tight.
\end{lemma}
\begin{Proof}
If neither of the constraints are tight, then the power policy $P(t)$ could be increased over a set of non-zero Lebesgue measure in the last epoch. This strictly increases the throughput, contradicting the optimality.
\end{Proof}

The following lemma shows that the optimal power should be monotonically decreasing between energy harvests.
\begin{lemma}
\label{new_lm}
$P(t)$ is piecewise monotone decreasing except possibly at the energy arrival instants. In particular, it is monotone decreasing between consecutive energy harvests.
\end{lemma}
\begin{Proof}
We prove the statement by contradiction. Assume that for some interval $[t_1,t_2]$, $P(t)$ is strictly monotone increasing, and that the interval $[t_1,t_2]$ does not contain an energy arrival instant. Define a new power policy as $P_{new}(t)=\frac{\int_{t_1}^{t_2}P(\tau)d\tau}{t_2-t_1}$ over $t \in [t_1,t_2]$ and $P_{new}(t)=P(t)$ otherwise. $P_{new}(t)$ satisfies the energy causality constraint in (\ref{opt_prob1}) since $P_{new}(t)$ uses the same amount of energy as $P(t)$ over $[t_1,t_2]$ and the energy constraint for $P(t)$ is not tight in this interval. $P_{new}(t)$ also satisfies the temperature constraint. To see this, we first note that $P_{new}(t)$ satisfies the following inequality (see \cite[Theorem on p. 207]{Ross06}):
\begin{align}\label{cnt1}
\int_{t_1}^{t_2} a e^{b\tau}P_{new}(\tau)d\tau \leq \int_{t_1}^{t_2} a e^{b\tau} P(\tau) d\tau
\end{align}
as both $P(t)$ and $e^{bt}$ are monotone increasing. In addition, since $P(t)$ is temperature feasible:
\begin{align}\label{cnt2}
\int_0^{t_1} a e^{b\tau}P(\tau)d\tau &\leq T_{\delta}e^{bt_1} \\
\int_{0}^{t_2} a e^{b\tau}P(\tau)d\tau &\leq T_{\delta}e^{bt_2} \label{cnt3}
\end{align}
Combining (\ref{cnt1}) and (\ref{cnt3}), we conclude that $P_{new}(t)$ satisfies the temperature constraint at $t=t_2$:
\begin{align}
\int_{0}^{t_2} ae^{b\tau}P_{new}(\tau)d\tau &=  \int_0^{t_1} ae^{b\tau}P_{new}(\tau)d\tau + \int_{t_1}^{t_2} ae^{b\tau}P_{new}(\tau)d\tau \\ &\leq \int_0^{t_1} ae^{b\tau}P(\tau)d\tau + \int_{t_1}^{t_2} ae^{b\tau}P(\tau)d\tau  \\ &\leq T_{\delta} e^{bt_2}
\end{align}
Additionally, the temperature constraint is satisfied for $t>t_2$ since $P_{new}(t)$ and $P(t)$ are identical for $t>t_2$ and $P(t)$ is temperature feasible. Hence, we need to show that $P_{new}(t)$ satisfies the temperature constraint for all $t \in (t_1,t_2)$ to establish the temperature feasibility of $P_{new}(t)$. That is, we need to show:
\begin{align}\label{gggg}
\int_0^{t_1} a e^{b\tau}P(\tau)d\tau + \int_{t_1}^{t} a e^{b\tau}P_{new}(\tau)d\tau \leq T_{\delta}e^{bt}, \quad t \in (t_1,t_2)
\end{align}
Since $P_{new}(t)=p$ is constant over $[t_1,t_2]$, we have:
\begin{align}\label{corin}
\int_{t_1}^{t} ae^{b\tau}P_{new}(\tau)d\tau = \frac{a}{b} p\left(e^{bt} - e^{bt_1}\right), \quad t \in [t_1,t_2]
\end{align}
Using (\ref{corin}) in (\ref{gggg}) and since $e^{bt} \geq 0$, (\ref{gggg}) takes the following equivalent form:
\begin{align} \label{ffgg}
e^{-bt} \left(\int_0^{t_1} ae^{b\tau}P(\tau)d\tau - \frac{a}{b}p e^{bt_1} \right) + \frac{a}{b}p \leq T_{\delta}
\end{align}
Note that the left hand side of (\ref{ffgg}) is either monotone increasing or monotone decreasing in $t$ as it is a linear function of $e^{-bt}$. Since the inequality (\ref{ffgg}) holds at $t=t_1$ and $t=t_2$ as $P_{new}(t)$ satisfies the temperature constraint at those points, we conclude that $P_{new}(t)$ satisfies the temperature constraint for all $t \in [t_1,t_2]$. In addition, $P_{new}(t)$ yields higher throughput than $P(t)$ due to the concavity of logarithm. This contradicts the optimality of $P(t)$. The proof holds even when $[t_1,t_2]$ includes an energy arrival instant provided that the energy causality constraint is not tight at that instant.
\end{Proof}

Next, we show that discontinuities in the power level could only occur in the form of positive jumps, and only at the instances of energy harvests.
\begin{lemma}\label{lm1}
If there is a discontinuity in $P(t)$, it is a positive jump and it occurs only at the energy arrival instants. The temperature $T(t)$ is continuous throughout the $[0,D]$ interval.
\end{lemma}
\begin{Proof}
Since $e^{bt}$ is a continuous function of $t$, $\lambda(t) \geq 0$ and $\beta(t)\geq 0$, any jump in $P(t)$ has to be positive due to (\ref{nn}). Any positive jump at instants other than $s_k$ violates monotonicity of $P(t)$ within each epoch due to Lemma~\ref{new_lm}. Due to (\ref{ths}), the resulting temperature $T(t)$ is continuous throughout the $[0,D]$ interval.
\end{Proof}

By Lemma~\ref{lm1}, we can take $\beta(t)$ in the form $\beta(t)=\sum_{j=1}^{N+1} \beta_j \delta(t-s_j)$ without loss of optimality, where $\beta_j \geq 0$, $j=1, \ldots, N+1$, are finitely many Lagrange multipliers corresponding to the energy causality constraints at the energy harvesting instants $s_j$ and the deadline, $s_{N+1}=D$.

The next lemma shows, for an arbitrary feasible policy $P(t)$, that if the temperature reaches the critical level $T_c$ at some $t_h$, then the power just before $t_h$ must be larger than a threshold.
\begin{lemma}\label{lm2}
If $T(t_h)=T_c$ for some $t_h \in [0,D)$, then $P(t_h-\epsilon) \geq \frac{T_{\delta}b}{a}$ for all sufficiently small $\epsilon>0$.
\end{lemma}
\begin{Proof}
Since $T(t_h)=T_c$, we have:
\begin{align}\label{kk}
\int_0^{t_h} a e^{b \tau} P(\tau) d \tau = T_{\delta}e^{bt_h} \end{align}
We combine (\ref{gg}) with (\ref{kk}) to get
\begin{align}\label{kkk}
\int_t^{t_h} a e^{b \tau} P(\tau) d \tau \geq T_{\delta} \left( e^{b t_h} - e^{bt} \right), \qquad \forall t \in [0,t_h]
\end{align}
which implies in view of the continuity of $P(t)$ (except for the finitely many energy arrival instants) proved in Lemma~\ref{lm1} that $P(t_h-\epsilon) \geq \frac{T_{\delta}b}{a}$ for all sufficiently small $\epsilon>0$.
\end{Proof}

We next state the continuity of the optimal power policy $P(t)$ at points when it hits the critical temperature $T_c$.
\begin{lemma}
\label{kkk}
If $T(t_h)=T_c$ for some $t_h \in [0,D)$ then $P(t)$ is continuous at $t_h$ and $P(t_h)=\frac{T_{\delta}b}{a}$.
\end{lemma}
\begin{Proof}
The proof follows from Lemma~\ref{constant} and Lemma~\ref{lm2} and the fact that negative jumps in $P(t)$ are not allowed due to Lemma~\ref{lm1}.
\end{Proof}

Next, we show that when the temperature hits the boundary $T_c$, it has to return to $T_c$.
\begin{lemma}
\label{extend}
Whenever $T(t_h)=T_c$ for some $t_h<D$, there exists $t>t_h$ such that $T(t)=T_c$.
\end{lemma}
\begin{Proof}
Assume that $T(t_h)=T_c$ for some $t_h<D$ and $T(t)<T_c$ for all $t_h<t<D$. By Lemma~\ref{kkk}, $P(t_h) = \frac{T_{\delta}b}{a}$. From (\ref{four}) with $T(t_h)=T_c$, the constraint $T(t) \leq T_c$ becomes:
\begin{align} \label{reef_r2}
\int_{t_h}^{t} a e^{b \tau} P(\tau)d\tau \leq  T_{\delta} \left( e^{b t} - e^{bt_h} \right), \quad t_h < t \leq D
\end{align}
Since $T(t)<T_c$ in $t_h<t<D$, only energy causality constraint is active and thus $P(t)$ for $t_h<t<D$ is the piecewise constant monotone power allocation in \cite{tcom-submit}. On the other hand, $P(t)=\frac{T_{\delta} b}{a}$ satisfies (\ref{reef_r2}) with equality for all $t$. Therefore, we must have $P(t) = c < \frac{T_{\delta}b}{a}$ for all $t \in (t_h,t_h+\delta)$ for some $\delta>0$. However, this contradicts $P(t_h) = \frac{T_{\delta}b}{a}$ since there cannot be a negative jump in $P(t)$ by Lemma~\ref{lm1}.
\end{Proof}

The following lemma identifies the exact conditions where the power $P(t)$ makes a jump.
\begin{lemma}
If there is a jump in $P(t)$, it occurs only at an energy arrival instant, when the battery is empty and the temperature is strictly below $T_c$.
\end{lemma}
\begin{Proof}
Due to the slackness conditions in (\ref{slck1})-(\ref{slck2}), a jump occurs if either the battery is empty or the temperature constraint is tight, i.e., $T(t)=T_c$. By Lemma~\ref{kkk}, $P(t)$ is continuous whenever $T(t)=T_c$. Therefore, a jump in $P(t)$ occurs at an energy arrival instant, when the battery is empty and $T(t)<T_c$.
\end{Proof}

We finally remark that energy may have to be wasted as aggressive use of energy may cause temperature to rise above the critical level.

\section{Optimal Policy in the Single Energy Arrival Case}

In this section, we consider a single epoch where $E$ units of energy is available at the transmitter at the beginning. We first develop further structural properties for the optimal power control policy in this specific case and then obtain the solution.

\subsection{Properties of an Optimal Policy}

The next lemma shows that, if the power falls below a certain threshold at an intermediate point and remains under that threshold until the deadline, then it should remain constant throughout.
\begin{lemma}
\label{lmn}
If $0< P(t) \leq \frac{T_{\delta}b}{a}$ for $t \in  [t_1,D]$, then $P(t)$ is constant over $[t_1,D]$.
\end{lemma}
\begin{Proof}
Assume $P(t)$ is not constant over $[t_1,D]$. Let $E_r=\int_{t_1}^{D} P(\tau) d\tau > 0$. Define a new policy $P_{new}(t) = \frac{E_r}{D - t_1}$ for $t \in [t_1,D]$ and $P_{new}(t)=P(t)$ otherwise. $P_{new}(t)$ is both energy and temperature feasible. Energy feasibility holds by construction as $P_{new}$ and $P$ have the same energy over $[t_1, D]$. Temperature feasibility also holds: $T(t_1)\leq T_c$ since $P(t)$ is temperature feasible and as $\frac{E_r}{D-t_1}<\frac{T_{\delta}b}{a}$, we have $T(t)\leq T_c$ for all $t_1<t<D$ from (\ref{gg}). Now, by Jensen's inequality $P_{new}(t)$ achieves strictly larger throughput since $\log$ is strictly concave. This contradicts the optimality of $P(t)$. Hence, $P(t)=c > 0$ for $t \in [t_1,D]$.\end{Proof}

The following lemma states that the power has to remain constant at the level $\frac{T_{\delta}b}{a}$ when the temperature reaches the critical level $T_c$.
\begin{lemma} \label{multiple_r}
Let $t' \in [0,D]$ denote $\min\{ t \in [0,D]: T(t)=T_c\}$. If $t' < D$, then $P(t)=\frac{T_{\delta}b}{a}$ for all $t \in [t',D]$.
\end{lemma}
\begin{Proof}
By Lemma~\ref{kkk}, $P(t')=\frac{T_{\delta}b}{a}$. By Lemma~\ref{new_lm}, $P(t)$ is monotone decreasing, and thus $0 \leq P(t) \leq \frac{T_{\delta}b}{a}$ for $t' < t \leq D$. By Lemma~\ref{lmn}, $P(t)=c$ for all $t \in [t',D]$. By Lemma~\ref{lm1}, $P(t)$ is continuous and therefore, $P(t)=\frac{T_{\delta}b}{a}$ for all $t \in [t',D]$.
\end{Proof}

The following lemma states that the optimal power is always larger than a constant value determined by the fixed system parameters.
\begin{lemma}\label{corl}
The optimal policy $P(t)$ satisfies:
\begin{align}
P(t) \geq \min\left\{\frac{T_{\delta}b}{a}, \frac{E}{D} \right\}, \quad \forall t \in [0,D]
\end{align}
\end{lemma}
\begin{Proof}
If the temperature constraint is not tight, then the problem reduces to the energy constrained problem in which case $P(t)=\frac{E}{D}$. If the temperature constraint is tight, $P(t)$ is monotone decreasing by Lemma~\ref{new_lm} and when the temperature level reaches $T_c$, $P(t)$ remains at $\frac{T_{\delta}b}{a}$ by Lemma~\ref{multiple_r}. Hence, $P(t) \geq \frac{T_{\delta}b}{a}$.
\end{Proof}

The following lemma shows that, since the power is always larger than a constant value, battery energy level is never zero, except possibly at the deadline.
\begin{lemma}\label{cor1}
In an optimal policy, energy in the battery is non-zero except possibly at $t=D$.
\end{lemma}
\begin{Proof}
By Lemma~\ref{corl}, the optimal power is always larger than a positive constant. Thus, the battery energy does not drop to zero.
\end{Proof}

The following lemma shows that the temperature is monotone increasing throughout the transmission duration, and also is a concave function of time.
\begin{lemma}\label{temperature}
The temperature with the optimal power policy is monotone increasing and concave.
\end{lemma}
\begin{Proof}
If the temperature constraint is never tight, then the optimal power level is $\frac{E}{D}$, and from Lemma~\ref{piecewise}, the temperature is monotone increasing. Concavity in this case follows from the concavity of the explicit expression in (\ref{fin}) with $T(0)=T_e$. Now, assume that the temperature constraint is tight at $t=D$. By Lemma~\ref{corl}, $P(t) \geq \frac{T_{\delta}b}{a}$. From (\ref{thermal}), we have:
\begin{align}\label{sxtn}
\frac{dT}{dt}&=aP(t) -b\left(T(t) - T_e\right) \\
&\geq a \frac{T_{\delta}b}{a} - b\left(T(t) - T_e\right) \\
&= b \left(T_c - T(t)\right) \geq 0 \label{svnt}
\end{align}
as $T(t) \leq T_c$ by the temperature constraint. Since $P(t)$ is monotone decreasing by Lemma~\ref{new_lm} and $T(t)$ is monotone increasing, from (\ref{sxtn}), $\frac{dT}{dt}$ is monotone decreasing, proving the concavity of $T(t)$ in this case.
\end{Proof}

\subsection{Optimal Policy}

In view of Lemma~\ref{cor1}, the energy constraint can be tight only at $t=D$. Therefore, the corresponding Lagrange multiplier is a single variable $\beta(t)=\beta \delta(t-D)$. From Lemma~\ref{temperature}, $T(t)$ is monotone increasing. Due to Lemma~\ref{multiple_r}, when $T(t)$ reaches $T_c$, power level has to remain at $\frac{T_{\delta}b}{a}$. Accordingly, we denote the instant when the temperature reaches $T_c$ as $t_0$.

\subsubsection{Sufficiently Large Energy}
In this case, the energy constraint is never tight, and $\beta=0$. In view of Lemma~\ref{lm12}, the temperature constraint is tight at $t=D$.

First, consider the case that $D$ is sufficiently large so that there exists $t_0<D$ such that $T(t_0)=T_c$. For $t \in [0,t_0)$, $T(t)<T_c$ and from (\ref{slck1}), $\lambda(t)=0$. From (\ref{nn}), when $t \in [0,t_0)$ we have $P(t)=\frac{1}{C}e^{-bt}-1$ where $C=\int_{t_0}^D\lambda(\tau)d\tau >0$. Since at $t=t_0$ the temperature reaches $T_c$, from Lemma~\ref{multiple_r}, we have $P(t)=\frac{T_{\delta}b}{a}$ for $t \in [t_0,D]$. Then, the optimal power has the form:
\begin{align}
P(t) = \left(\frac{1}{C}e^{-bt}-1\right)\left(u(t) - u(t-t_0)\right) + \frac{T_{\delta} b}{a}u(t-t_0) \label{kkl}
\end{align}
where $u(t)$ is the unit step function. Now, from Lemma~\ref{kkk}, $P(t)$ is continuous at $t_0$ and $C$ should be chosen accordingly. In particular, $C=\frac{1}{\left(\frac{T_{\delta} b}{a}+1\right)}e^{-bt_0}$. The following Lagrange multiplier $\lambda(t)$ verifies (\ref{kkl}):
\begin{align}
\lambda(t) = \frac{b}{\left(\frac{T_{\delta}b}{a} + 1\right)}e^{-bt}u(t-t_0) + \frac{e^{-bD}}{\left(\frac{T_{\delta}b}{a}+1\right)} \delta(t-D)
\end{align}
The corresponding optimal temperature pattern for $0 \leq t \leq t_0$ is:
\begin{align}
T(t)=a\left(\frac{T_{\delta}b}{a} + 1\right)te^{-b(t-t_0)} + \frac{a}{b}e^{-bt} - \frac{a}{b} + T_e
\end{align}
and $T(t)=T_c$ for $t_0 \leq t \leq D$. We note that $t_0$ satisfies:
\begin{align}\label{sol}
\left(\frac{T_{\delta}}{a} + \frac{1}{b}\right)e^{bt_0} - \frac{1}{b}=\left(\frac{T_{\delta}b}{a} + 1\right)t_0e^{bt_0}
\end{align}
so that $T(t_0)=T_c$. Hence, $T(t)$ monotonically increases till it reaches $T_c$, which is consistent with Lemma~\ref{temperature}.

Next, consider the case that $D<t_0$. In this case,
\begin{align}\label{dld}
P(t)=\frac{1}{C} e^{-bt} - 1
\end{align}
where $C = \frac{D}{\left(\left(\frac{T_{\delta}}{a}+\frac{1}{b}\right)e^{bD} - \frac{1}{b}\right)}$ and $\lambda(t) = C\delta(t-D)$. Therefore, the optimal $P(t)$ in this case is
\begin{align}\label{halfway}
P(t)=\frac{1}{D}\left(\left(\frac{T_{\delta}}{a}+\frac{1}{b}\right)e^{bD} - \frac{1}{b}\right)e^{-bt} -1
\end{align}

We also remark that $t_0$ level that satisfies (\ref{sol}) monotonically increases with $T_{\delta}$. To see this, we rearrange (\ref{sol}) as follows:
\begin{align}\label{sol2}
\frac{1}{b}\left(1-\frac{1}{\left(\frac{T_{\delta}b}{a} + 1\right)}e^{-bt_0}\right)-t_0=0
\end{align}
Let us define a multi-variable real function $w(t_0,T_{\delta})$ as the left hand side of (\ref{sol2}) and denote a specific solution as $t_0^{*}$ for fixed $T_{\delta}$. It is easy to see that (\ref{sol2}) always has a solution $t_0$ for fixed $T_{\delta}$. To see this, we evaluate the derivative with respect to $t_0$ as:
\begin{align}
\frac{\partial}{\partial t_0} w(t_0,T_{\delta})&=\frac{1}{\left(\frac{T_{\delta}b}{a} + 1\right)}e^{-bt_0}-1 \leq 0, \quad \forall t_0 \geq 0
\end{align}
That is, $w(t_0,T_{\delta})$ is monotone decreasing with $t_0$. At $t_0=0$, $w(t_0,T_{\delta})>0$ while $w(t_0,T_{\delta})\rightarrow - \infty$ as $t_0$ grows. In view of the continuity of $w(t_0,T_{\delta})$, there exists a $t_0$ such that $w(t_0,T_{\delta})=0$. Additionally, we observe in (\ref{sol2}) that for fixed $t_0$, $w(t_0,T_{\delta})$ monotonically increases with $T_{\delta}$. Therefore, if $w(t_0^*,T_{\delta})=0$, then, due to monotone increasing property with respect to $T_{\delta}$, $w(t_0^*,T_{\delta}^{'})>0$ for $T_{\delta}^{'}>T_{\delta}$. Hence, for $t_0^{**}$ such that $w(t_0^{**},T_{\delta}^{'})=0$, we have $t_0^{**}>t_0^{*}$ due to monotone decreasing property with respect to $t_0$.

\subsubsection{Energy Limited Case}
Note that the optimal power policies in the energy unconstrained cases in (\ref{kkl}) and (\ref{halfway}) have finite energies. If the available energy $E$ is larger than the corresponding energy level in (\ref{kkl}) and (\ref{halfway}), then the solution is as in (\ref{kkl}) and (\ref{halfway}). Otherwise, the energy constraint is active and the Lagrange multiplier is  $\beta > 0$. From (\ref{nn}), we have:
\begin{align}
P(t) = \frac{1}{\beta + e^{bt}\int_t^D \lambda(\tau)d\tau} - 1
\end{align}

We first note that there is a critical energy level $E_{critical}$ such that if $E\leq E_{critical}$, then constant power policy $P(t)=\frac{E}{D}$ is optimal. This critical level is:
\begin{align}\label{thrtn}
E_{critical} = \frac{T_{\delta}b}{a}\frac{De^{bD}}{e^{bD} - 1}
\end{align}
This is the critical level below which the temperature constraint is not tight by the constant power allocation $P(t)=\frac{E}{D}$. The expression in (\ref{thrtn}) is evaluated from (\ref{fin}) by inserting $T(0)=T_e$, and requiring $T(D) \leq T_c$. When $E \leq E_{critical}$, $\lambda(t)=0$ since temperature constraint is never tight. In this case, $\beta = \frac{1}{\frac{E}{D}+1}$. $E_{critical}$ is the maximum energy level for which a constant power level is optimal. If $P(t)=\frac{E_{critical}}{D}$, $T(t)$ is monotone increasing over $[0,D]$ and reaches $T_c$ at $t=D$. If $E>E_{critical}$, the constant power level $\frac{E_{critical}}{D}$ does not satisfy the temperature constraint. We note from (\ref{thrtn}) that $E_{critical}$ increases with the deadline $D$. Therefore, there exists a deadline level $\tilde{D}$ for which $D>\tilde{D}$ implies $E<E_{critical}$ and hence constant power policy is optimal.

An alternative way of observing the behavior of the optimal policy is to fix the available energy $E$ and $T_e$ and vary the critical temperature $T_c$. In this case, there is a critical temperature limit $T_c^{limit}$ for which $P(t)=\frac{E}{D}$ is optimal whenever $T_c>T_c^{limit}$:
\begin{align}
T_c^{limit}=T_e + \frac{a}{b}\frac{E}{D}\frac{e^{bD} - 1}{e^{bD}}
\end{align}
which again is evaluated from (\ref{fin}) with $T(0)=T_e$. In the following, we consider $E>E_{critical}$ or $T_c < T_c^{limit}$ so that both energy and temperature constraints are tight at the end of the communication session.

Again, we consider two possibilities: temperature constraint becomes tight at a $t_0<D$, and temperature constraint becomes tight at $t=D$. In both cases, the energy constraint becomes tight at $t=D$.

First, consider the case that $t_0<D$: Due to (\ref{slck1}), $\lambda(t)=0$ for $t \in [0,t_0)$ and from (\ref{nn}), we get:
\begin{align}\label{yyy}
P(t)=\frac{1}{\beta + Ce^{bt}} - 1
\end{align}
where $C=\int_{t_0}^D\lambda(\tau)d\tau >0$. Additionally, $P(t)=\frac{T_{\delta}b}{a}$ for the remaining portion of the epoch in view of Lemma~\ref{multiple_r}. $t_0$ is such that for $t>t_0$, $P(t)=\frac{T_{\delta}b}{a}$ and $T(t_0)=T_c$. Since $P(t_0)=\frac{T_{\delta}b}{a}$ we have:
\begin{align}\label{bir}
\frac{1}{\beta + Ce^{bt_0}} = \frac{T_{\delta}b}{a} + 1
\end{align}
Similarly, for $T(t_0)=T_c$, we have from (\ref{ths}) with $T(0)=T_e$:
\begin{align}
e^{-bt_0} \left(\int_0^{t_0} e^{bt} \left(a\left(\frac{1}{\beta + Ce^{bt}} - 1\right) + bT_e\right) dt + T_e \right)=T_c
\end{align}
Finally, the energy constraint has to be satisfied at $t=D$:
\begin{align}
\int_0^{t_0} \left(\frac{1}{\beta + Ce^{bt}} - 1\right)dt + \frac{T_{\delta}b}{a}\left(D-t_0\right)= E \label{uc}
\end{align}
If there exists $t_0 \leq D$ for (\ref{bir})-(\ref{uc}), then $P(t)$ is:
\begin{align}\nonumber
P(t) = &\left( \frac{1}{\beta + Ce^{bt}} - 1 \right)\left(u(t) - u(t-t_0)\right) + \frac{T_{\delta} b}{a}u(t-t_0)
\end{align}
In this case, the corresponding Lagrange multiplier is:
\begin{align}
\lambda(t) = bCe^{-b(t-t_0)}u(t-t_0) + Ce^{b(t_0-D)}\delta(t-D)
\end{align}

Otherwise, when no such $t_0<D$ exists, the temperature constraint is tight only at $t_0=D$. In this case, $P(t)$ is as in (\ref{yyy}) for $t \in [0,D]$ where $\beta$ and $C$ have to satisfy:
\begin{align}
e^{-bD} \left(\int_0^{D} e^{bt} \left(a\left(\frac{1}{\beta + Ce^{bt}} - 1\right) + bT_e\right) dt + T_e \right)&=T_c
\\ \int_0^{D} \left(\frac{1}{\beta + Ce^{bt}} - 1\right)dt &= E
\end{align}
The corresponding Lagrange multiplier is $\lambda(t) = C\delta(t-D)$.

\begin{figure}[t]
\centering
\subfigure[Fixed $T_c$ and varying $E$.]{\label{reg_o1}
\includegraphics[width=0.49\linewidth]{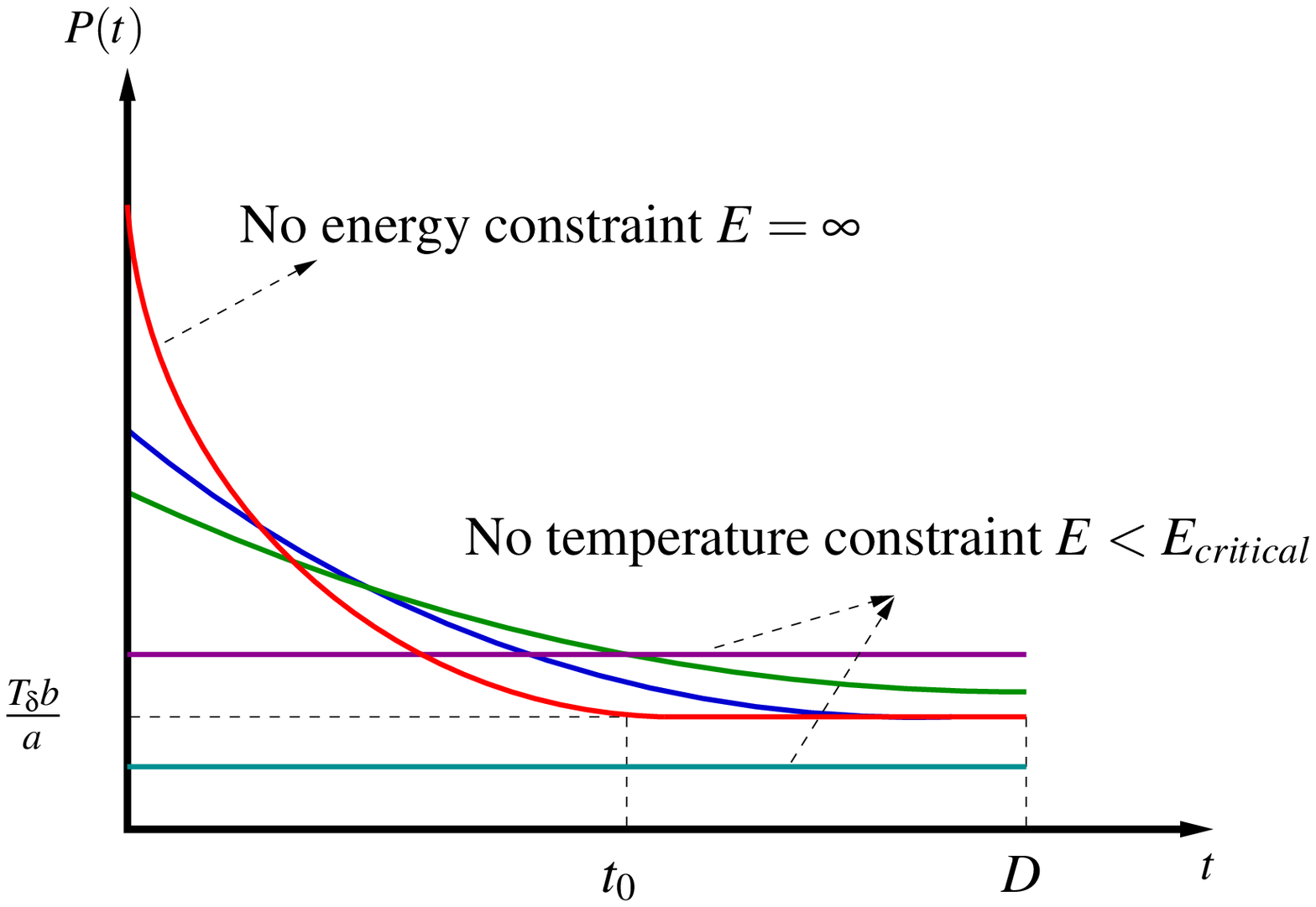}}
\subfigure[Fixed $E$ and varying $T_c$.]{\label{reg_o2}\hspace{-0.15in}
\includegraphics[width=0.50\linewidth]{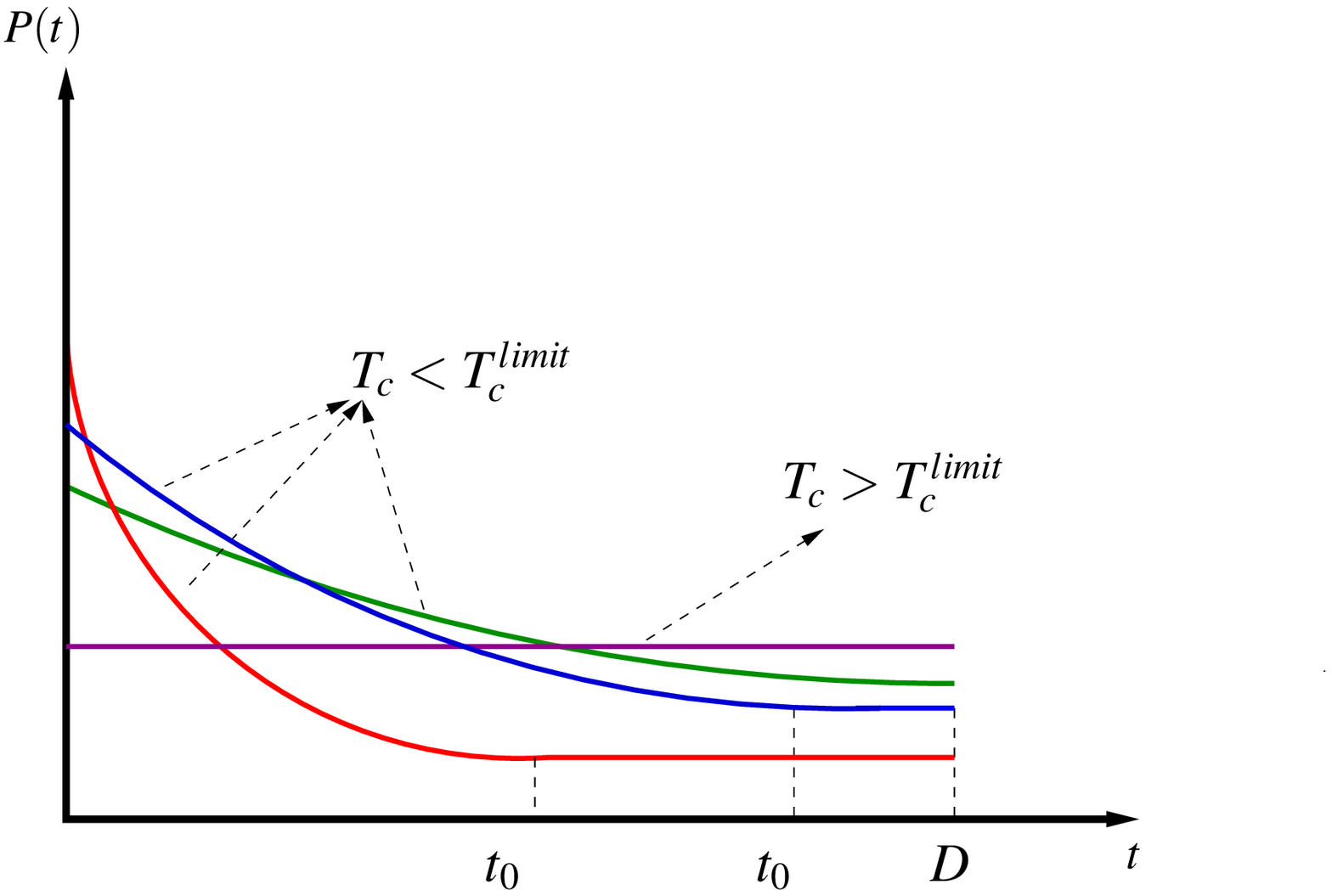}}
\caption{The optimal power policy in the single energy arrival case for different energy and deadline constraints.}
\label{reg_o}
\end{figure}

Depending on the energy $E$ and the critical temperature $T_c$, the optimal power scheduling policy $P(t)$ varies according to the plots in Fig.~\ref{reg_o}. For small $E$ and fixed $T_c$ or for large $T_c$ and fixed $E$, a constant power policy is optimal. For moderate and large $E$, the optimal power policy is exponentially decreasing and may hit the power level $\frac{T_{\delta} b}{a}$. Note that $t_0$ level at which temperature touches the critical level decreases as $T_c$ is decreased and as $E$ is increased. In particular, for fixed $T_c$, the level of $t_0$ is bounded below by the solution for $E=\infty$ whereas for fixed $E$, $t_0$ goes to $0$ as $T_c$ approaches $T_e$.

\section{Optimal Policy for Multiple Energy Arrivals}

In this section, we extend the solution to the case of multiple energy arrivals. We start with extending the properties observed for the single energy arrival case when initial temperature $T(0)$ is different from $T_e$. The following lemma generalizes Lemmas~\ref{new_lm}, \ref{multiple_r} and \ref{temperature} for the case of an arbitrary $T(0)$.

\begin{lemma} \label{extension}
Assume that the initial temperature $T(0)$ is in the range $T_e < T(0) < T_c$ instead of $T(0)=T_e$ and consider the single energy arrival case: $P(t)$ is monotone decreasing. Let $t_h \in [0,D]$ denote $\min\{ t \in [0,D]: T(t)=T_c\}$. If $t_h < D$, then $P(t)=\frac{T_{\delta}b}{a}$ for all $t \in [t_h,D]$ and the temperature is monotone increasing and concave. If $T(0)=T_c$, then $P(t)=\min\left\{\frac{T_{\delta}b}{a},\frac{E}{D}\right\}$.
\end{lemma}
\begin{Proof}
If $T(0)$ is in the range $T_e<T(0) < T_c$ then, instead of (\ref{gg}), we have the following temperature constraint:
\begin{align} \label{gg2}
\int_0^{t} a e^{b \tau} P(\tau) d \tau \leq T_{\delta}e^{bt} - T_g, \quad \forall t \in [0,D]
\end{align}
where $T_g = T(0) - T_e\geq 0$. Note that $T_{\delta}e^{bt} - T_g \geq 0$ for all $t \in [0,D]$, i.e., the right hand side of (\ref{gg2}) is always non-negative. The argument in Lemma~\ref{new_lm} is valid in the presence of the additional term $T_g$ in (\ref{gg2}), and therefore $P(t)$ is monotone decreasing.

The second claim follows from the argument in Lemma~\ref{multiple_r}. In particular, in addition to Lemma~\ref{new_lm}, Lemma~\ref{lmn} directly extends with the constraint in (\ref{gg2}). Hence, the result follows by applying the argument in Lemma~\ref{multiple_r}.

Finally, $T(t)$ is monotone increasing and concave due to the steps followed in Lemma~\ref{temperature}. In particular, if the temperature constraint is tight at $t=D$, $P(t) \geq \frac{T_{\delta}b}{a}$. Hence, (\ref{sxtn})-(\ref{svnt}) hold and the temperature is monotone increasing and concave. If $T(0)=T_c$, then $P(t)=\min\left\{\frac{T_{\delta}b}{a},\frac{E}{D}\right\}$ due to the energy constraint. Note that the temperature decreases in case $\min\left\{\frac{T_{\delta}b}{a},\frac{E}{D}\right\}=\frac{E}{D}$.
\end{Proof}

As in the single epoch case, we will investigate the solution under special cases. In particular, we will investigate the solution according to the time when the temperature hits the critical level. To this end, we specialize in an interval $[t_1,t_2]$ such that $T(t)<T_c$ for all $t \in [t_1,t_2)$ and $T(t_2)=T_c$ where $0<t_1 < t_2 \leq D$. Note that the temperature $T(t)$ is a continuous function of $t$ and hence there exist such intervals. We assume that the solution is known in $[0,t_1) \cup (t_2,D]$ and we let $T_e \leq T(t_1) < T_c$. In this case, the solution of (\ref{opt_prob1}) over the interval $[t_1,t_2]$ is equal to the solution of the following problem obtained by restricting the temperature constraint to be satisfied at $t=t_2$ only:
\begin{align}\nonumber
\max_{P(t), \ t \in [t_1,t_2]} \quad &\int_{t_1}^{t_2} \frac{1}{2}\log\left(1 + P(\tau)\right)d\tau \\ \nonumber
\mbox{s.t.} \quad &\int_{t_1}^{t_2} a e^{b \tau} P(\tau) d \tau = T_{\delta}e^{bt_2} - T_g \\
\qquad &\int_{0}^{t} P(\tau) d\tau \leq \sum_{i=0}^{\tilde{h}(t)} \tilde{E}_i, \qquad \forall t \in [t_1,t_2] \label{opt_prob_relax}
\end{align}
where $T_g=T(t_1)-T_e \geq 0$. In (\ref{opt_prob_relax}), $\tilde{E}_i$ is determined as follows: $\tilde{E}_0$ is the available energy in the battery at time $t=t_1$. $\tilde{E}_i$ for $i=1,\ldots,\tilde{N}$ are the energy arrivals at instants $\tilde{s}_i \in (t_1,t_2)$. $\tilde{h}(t)$ is defined accordingly. While the times $\tilde{s}_i$ are exactly those in the original problem, the amounts $\tilde{E}_i$ may be different from the original amounts as some energy may be left for use in the $(t_2,D]$ interval. For the following argument, whether $\tilde{E}_i$ equals the original energy arrival amount is not relevant and we leave $\tilde{E}_i$ as arbitrary amounts. To obtain the solution of (\ref{opt_prob_relax}) using this Lagrangian framework, it is necessary and sufficient to find $\tilde{N}+2$ variables $\beta_i \geq 0$, $i=1, \ldots, \tilde{N}+1$ and $C \geq 0$ such that
\begin{align}\label{pow}
P(t)=\left[ \frac{1}{\sum_{j=i}^{\tilde{N}+1}\beta_j + Ce^{bt}} - 1\right]^+, \quad t \in [\tilde{s}_{i-1},\tilde{s}_{i}), \ i=1,\ldots, \tilde{N}+1
\end{align}
with the corresponding slackness conditions. Therefore, for the $[t_1,t_2]$ interval, the solution has the structure in (\ref{pow}), which is parameterized by finitely many Lagrange multipliers. In particular, throughout an epoch over which $T(t)<T_c$, power level satisfies $P(t)=\left[ \frac{1}{\beta+Ce^{bt}} - 1 \right]^{+}$ for some $\beta \geq 0$ and $C \geq 0$ not both equal to zero. This also holds in a subinterval of an epoch over which $T(t)<T_c$. In the following lemma, we show that in such an epoch, the temperature $T(t)$ is unimodal.

\begin{lemma}
\label{unimodal}
If $P(t)=\left[ \frac{1}{\beta+Ce^{bt}}-1 \right]^{+}$ for $t \in [t_1,t_2]$ for some $\beta >0$ and $C>0$, the resulting $T(t)$ is unimodal over $[t_1,t_2]$.
\end{lemma}
\begin{Proof}
From (\ref{four}), we have for $t \in [t_1,t_2]$,
\begin{align}\label{temp_exp}
T(t)= e^{-b(t-{t_1})} \left(\int_{t_1}^{t} e^{b(\tau-t_1)} \left(a\left[\frac{1}{\beta + Ce^{b\tau}} - 1\right]^+ + bT_e\right) d \tau + T(t_1) \right)
\end{align}
First, we note that when $P(t)=0$, $\frac{d}{dt}T(t) \leq 0$ from (\ref{thermal}). Hence, it suffices to show that $T(t)$ is unimodal when $P(t)=\frac{1}{\beta+Ce^{bt}}-1 > 0$. By evaluating the integral, we get
\begin{align}\label{lbbb}
T(t)=\frac{a}{bC}e^{-bt}\log\left(\frac{\beta+Ce^{bt}}{\beta+Ce^{bt_1}}\right)+\left(T(t_1)-T_e+\frac{a}{b}\right)e^{-b(t-t_1)} + T_e - \frac{a}{b}
\end{align}
We claim that $T(t)$ in (\ref{lbbb}) is unimodal for $t>t_1$. Note that the derivative of $T(t)$ is:
\begin{align}\label{dfsa}
\frac{d}{dt}T(t)=e^{-bt}\left( \frac{ae^{bt}}{\beta + Ce^{bt}} -  \frac{a}{C}\log\left(\frac{\beta+Ce^{bt}}{\beta+Ce^{bt_1}} \right) - b \left(T(t_1)-T_e+\frac{a}{b}\right)e^{bt_1} \right)
\end{align}
We let $x=e^{bt}$, $x_1=e^{bt_1}$ and concentrate on $\frac{ax}{\beta + Cx} -  \frac{a}{C}\log\left(\frac{\beta+Cx}{\beta+Cx_1} \right)$ for $x>x_1$. We note that $\frac{ax}{\beta + Cx} -  \frac{a}{C}\log\left(\frac{\beta+Cx}{\beta+Cx_1} \right)$ is a strictly monotone decreasing function of $x$ for $x>x_1>0$. In particular, we have:
\begin{align}
\frac{d}{dx}\left( \frac{ax}{\beta + Cx} -  \frac{a}{C}\log\left(\frac{\beta+Cx}{\beta+Cx_1} \right) \right) = \frac{-Cx}{\left(\beta + Cx\right)^2}
\end{align}
Thus, $\frac{ae^{bt}}{\beta + Ce^{bt}} -  \frac{a}{C}\log\left(\frac{\beta+Ce^{bt}}{\beta+Ce^{bt_1}} \right)$ is strictly monotone decreasing in $t$. As $\frac{ae^{bt}}{\beta + Ce^{bt}} -  \frac{a}{C}\log\left(\frac{\beta+Ce^{bt}}{\beta+Ce^{bt_1}} \right) >0$ at $t=t_1$, we conclude that the factor in (\ref{dfsa}) that multiplies $e^{-bt}$ can take value $0$ at most once. In particular, $\frac{ae^{bt}}{\beta + Ce^{bt}} -  \frac{a}{C}\log\left(\frac{\beta+Ce^{bt}}{\beta+Ce^{bt_1}} \right) - b \left(T(t_1)-T_e+\frac{a}{b}\right)e^{bt_1}$ can take positive or negative values at $t=t_1$. If it is positive at $t=t_1$, it hits value $0$ at most once for $t>t_1$. If it is negative at $t=t_1$, it stays negative throughout $t>t_1$. This proves that $T(t)$ is unimodal over $[t_1,t_2]$.
\end{Proof}

In the following lemma, we show that, in an epoch $[s_i,s_{i+1}]$, the temperature cannot return to $T_c$ if it hits and falls below $T_c$.

\begin{lemma}
\label{new}
If $T(t_h)=T_c$ and $T(t_h+\Delta)<T_c$ for some $\Delta>0$ where both $t_h$ and $t_h+\Delta$ are in $[s_i,s_{i+1}]$, then $T(t)<T_c$ for all $t \in [t_h+\Delta,s_{i+1}]$.
\end{lemma}
\begin{Proof}
By Lemma~\ref{kkk}, $P(t_h)=\frac{T_{\delta}b}{a}$. By Lemma~\ref{new_lm}, power is monotone decreasing in an epoch. Therefore, if $T(t_h+\Delta)<T_c$, then $P(t_h+\Delta)<\frac{T_{\delta}b}{a}$ and hence $P(t)<\frac{T_{\delta}b}{a}$ for all $t \in [t_h+\Delta,s_{i+1}]$. This, in turn, means that $T(t)<T_c$ for all $t \in [t_h+\Delta,s_{i+1}]$.
\end{Proof}

Next, we complete the unimodal structure of the temperature by showing that it has to be monotone decreasing if it hits and falls below $T_c$.
\begin{lemma}\label{new2}
In an epoch $[s_i,s_{i+1}]$, if the temperature touches $T_c$ at $t_h$ and falls below it, then the temperature is monotone decreasing in $[t_h,s_{i+1}]$.
\end{lemma}
\begin{Proof}
By Lemma~\ref{new}, if $T(t_h+\Delta)<T_c$, then $T(t)<T_c$ for all $t \in [t_h+\Delta,s_i]$. Therefore, we have
\begin{align}
P(t)= \left[\frac{1}{\beta+Ce^{bt}} - 1 \right]^{+}, \quad t \in [t_h+\Delta,s_{i+1}]
\end{align}
for some $\beta>0$ and $C>0$. By Lemma~\ref{unimodal}, $T(t)$ is unimodal over $t \in [t_h+\Delta,s_i]$. Therefore, $T(t)$ is monotone decreasing.
\end{Proof}

We next consider epochs $[s_i,s_{i+1}]$ and its subintervals over which $T(t)<T_c$ and $T(t)=T_c$. By Lemma~\ref{unimodal} and in view of the discussion around (\ref{opt_prob_relax}), whenever $T(t)<T_c$ over an epoch, $T(t)$ reaches its peak level over that epoch at only one instance. Consequently, if $T(t)<T_c$ for all $t \in [s_i,s_{i+1}]$, there are three possible cases. The first two possibilities are that $T(t)$ is monotone increasing or monotone decreasing throughout the epoch. The third possible case is that $T(t)$ is monotone increasing in $[s_i,t_{1i}]$ and monotone decreasing in $(t_{1i},s_{i+1}]$ for some $t_{1i} \in (s_i,s_{i+1})$. Otherwise, $T(t)$ hits $T_c$ and $T(t)$ does not return to $T_c$ if it falls below it due to Lemma~\ref{new}. Therefore, if $T(t)$ hits $T_c$ in an epoch $[s_i, s_{i+1}]$, then that epoch is divided into three successive subintervals $I_{i1}, I_{i2}, I_{i3}$ with $I_{i1}=[s_i,t_{i1})$, $I_{i2}=[t_{i1},t_{i2})$ and $I_{i3}=[t_{i2},s_{i+1}]$ for some $s_{i}<t_{i1}\leq t_{i2} < s_{i+1}$. $T(t)$ is monotone increasing over $I_{i1}$, remains at $T_c$ over $I_{i2}$ and is monotone decreasing over $I_{i3}$. We finally note that if $T(t)<T_c$ at $t=D$, then $T(t)<T_c$ for all $t \in [0,D]$. This follows from Lemma~\ref{extend}. In this case, the temperature constraint is never tight and the optimal power policy is identical to the one in \cite{tcom-submit}.

\begin{figure}[t]
\centerline{\includegraphics[width=0.85\linewidth]{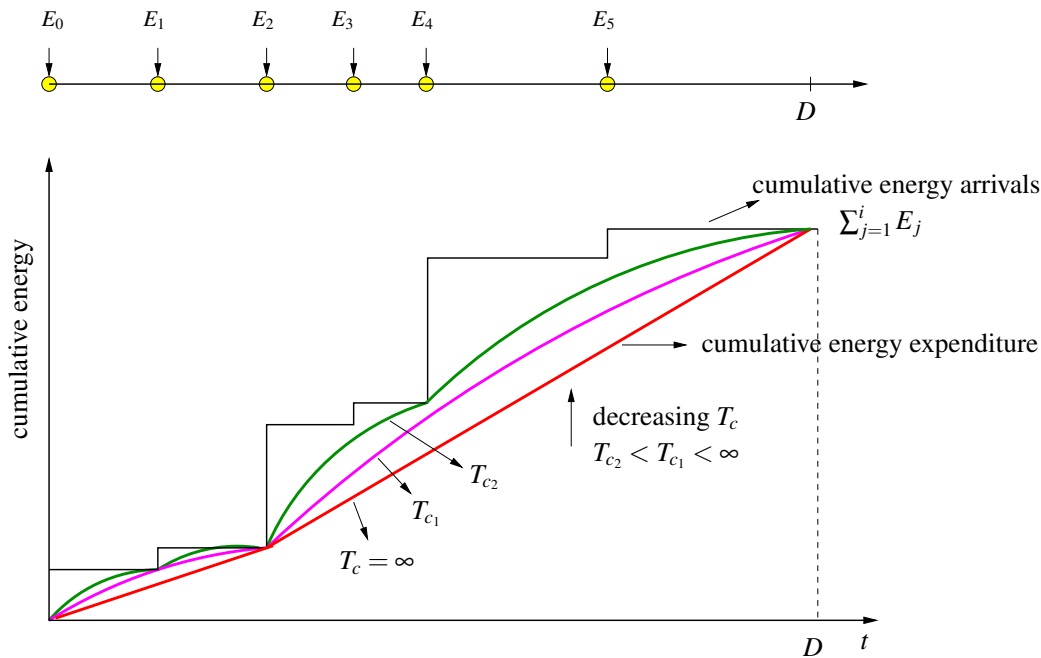}}
\caption{Energy expenditure with the optimal power policy with multiple energy arrivals. In view of the temperature constraint, as $T_c$ is decreased, the energy is spent faster subject to energy causality. }
\label{exp2}
\end{figure}

In Fig.~\ref{exp2}, we plot the optimal energy expenditure for different values of critical temperature level $T_c$. We observe that as $T_c$ is decreased, the temperature budget shrinks and the temperature constraint becomes more likely to be tight. In this case, energy is spent faster not to create high amounts of heat in the system. In general, there is a tension between causing unnecessary heat in the system and maximizing the throughput. While we have fully characterized this tension in the single energy arrival case, it needs to be further explored in the multiple energy arrivals case. In particular, when a high amount of energy arrives into the system during the progression of communication, the transmitter has to accommodate it by cooling down and creating a temperature margin for future use. While maximizing the throughput generally requires using the energy in the system to the fullest extent, the transmitter may have to waste energy due to the temperature limit. We investigate this tension in numerical examples in the next section.

\section{Numerical Results}

\begin{figure}[t]
\centerline{\includegraphics[width=0.85\linewidth]{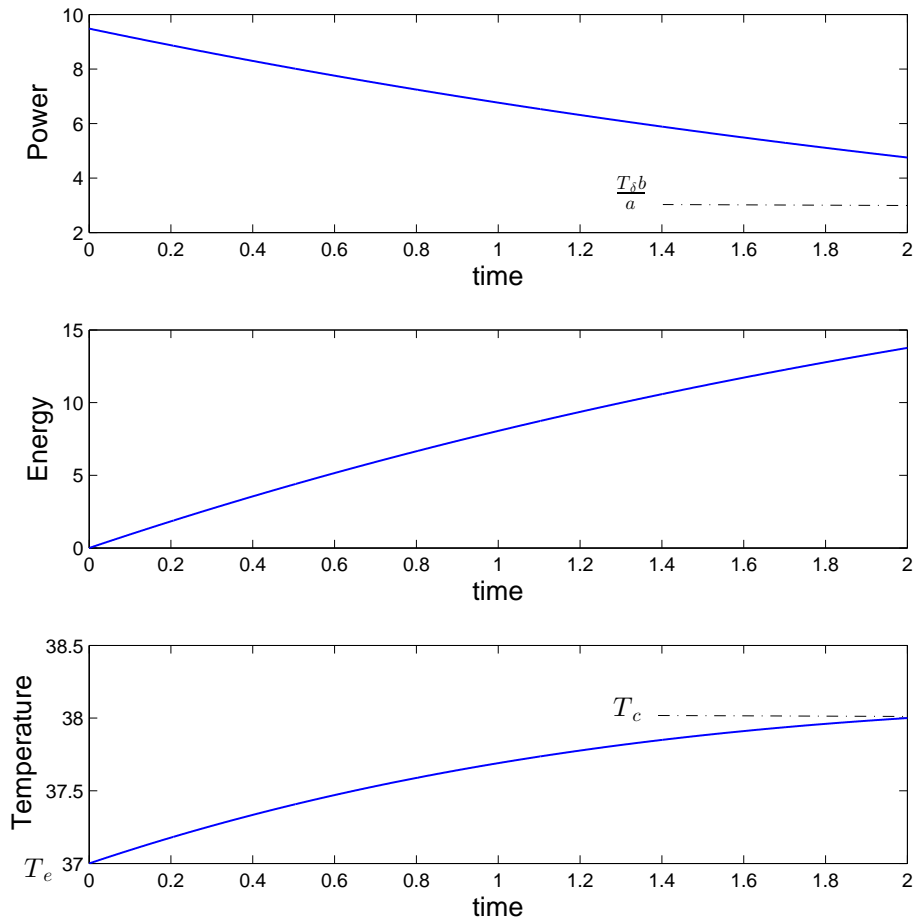}}\vspace{-0.3in}
\caption{Power, energy and temperature plots for unlimited energy and $D = 2$ for the single epoch case.}
\label{model4}
\end{figure}

\begin{figure}[t]
\centerline{\includegraphics[width=0.85\linewidth]{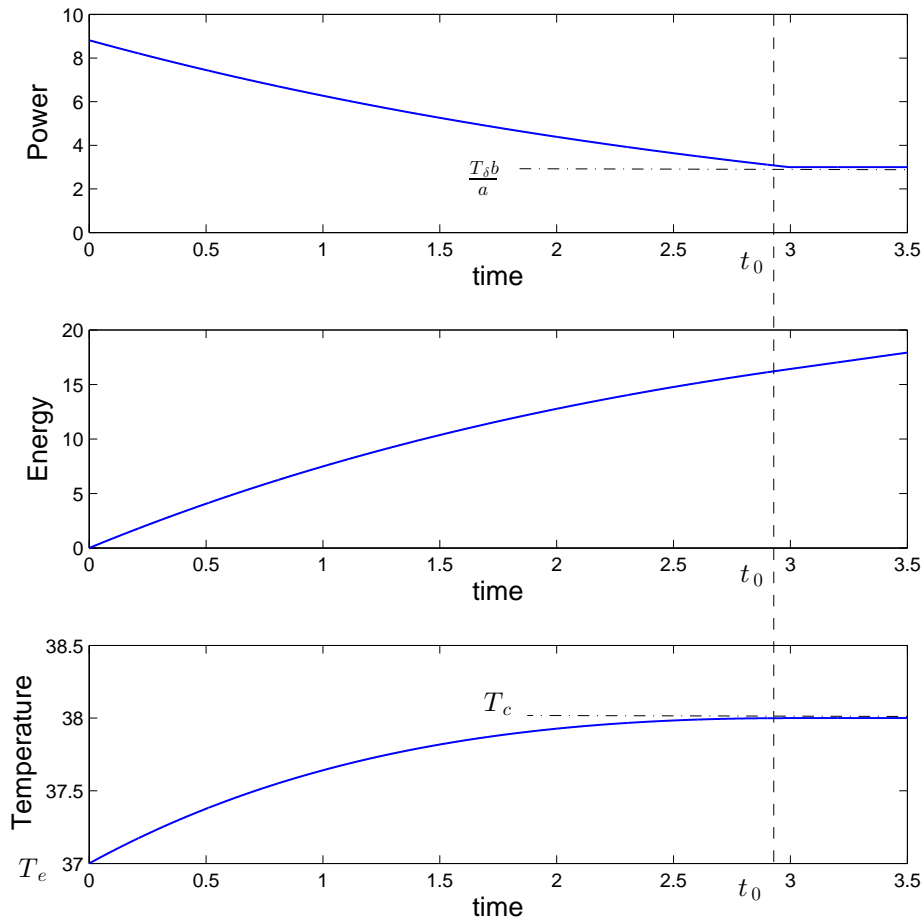}}\vspace{-0.3in}
\caption{Power, energy and temperature plots for unlimited energy and $D = 3.5$ for the single epoch case.}
\label{model5}
\end{figure}

\begin{figure}[t]
\centerline{\includegraphics[width=0.85\linewidth]{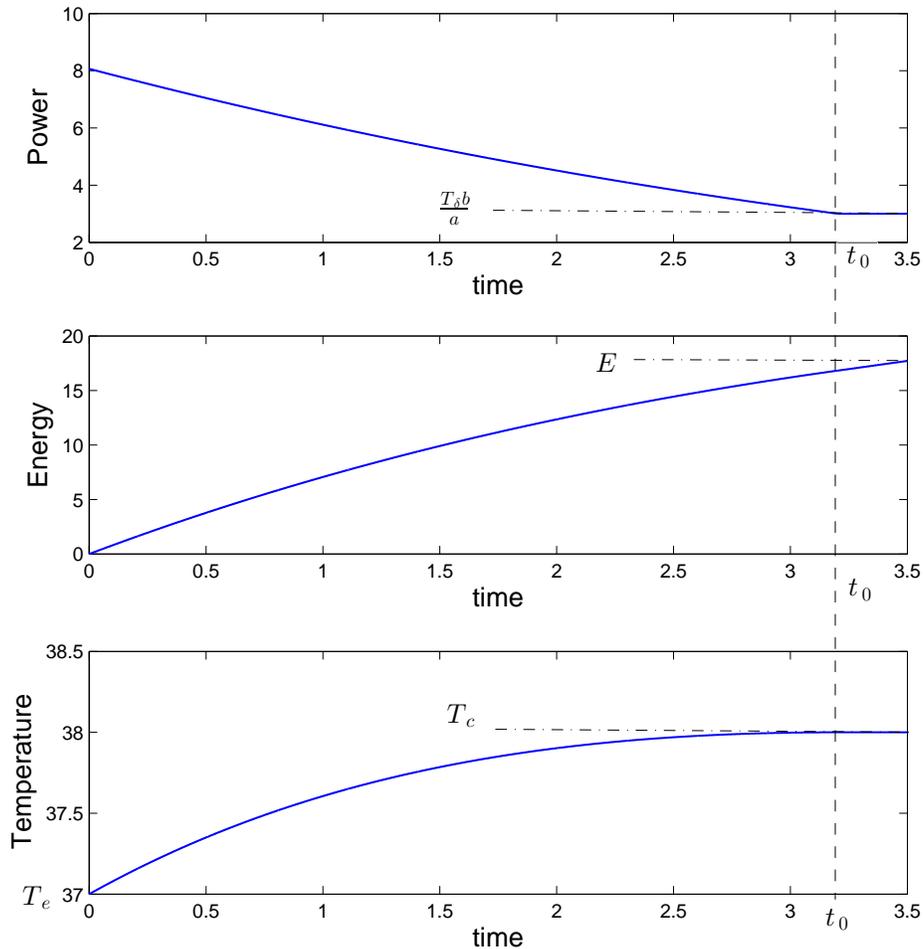}}\vspace{-0.3in}
\caption{Power, energy and temperature plots for limited energy $E=17.71$ and $D = 2$ for the single epoch case.}
\label{model6}
\end{figure}

\begin{figure}[t]
\centerline{\includegraphics[width=0.87\linewidth]{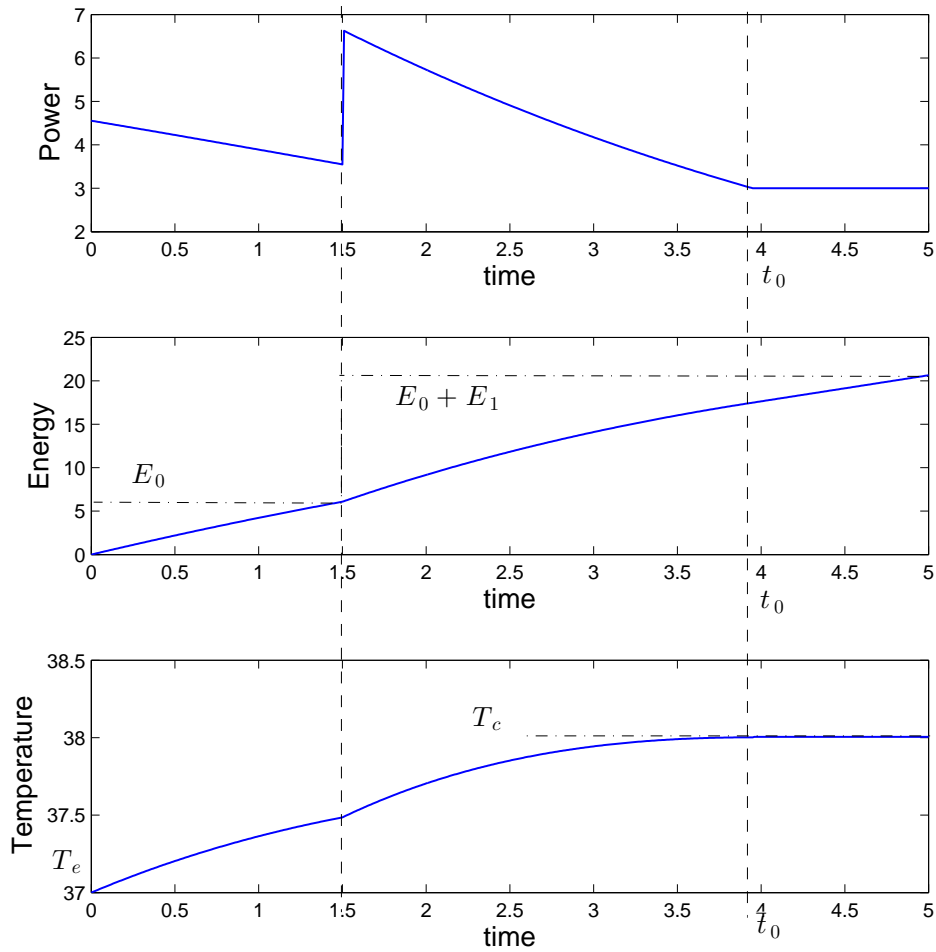}}\vspace{-0.3in}
\caption{Power, energy and temperature plots for two energy arrivals, $E_0=6.08$ and $E_1=14.55$ at $t=1.5$ and $D=5$.}
\label{model7}
\end{figure}

In this section, we provide numerical examples to illustrate the optimal power policy and the resulting temperature profile. For plots in Figs. \ref{model4}, \ref{model5}, \ref{model6} and \ref{model7}, we set $a=0.1$, $b=0.3$, $T_{e}=37$ and $T_{c}=38$. Therefore, the critical power level is $\frac{T_{\delta}b}{a}=3$.

In Figs. \ref{model4} and \ref{model5}, we consider the energy unlimited scenario. In this case, the solution of (\ref{sol}) is found as $t_0=2.993$. In Fig.~\ref{model4}, we set $D=2 < t_0$ and we observe that the optimal power policy is always above the level $\frac{T_{\delta}b}{a}$. In this case, power strictly monotonically decreases while temperature strictly monotonically increases with temperature touching the critical level $T_c$ at the deadline. In Fig.~\ref{model5}, we set the deadline as $D=3.5 > t_0$. We calculate that the energy needed to have the power policy in Fig.~\ref{model5} is $E=17.98$. In other words, if the initial energy is $E \geq 17.98$ then the power policy in Fig.~\ref{model5} is optimal. We observe that the optimal power level monotonically decreases to the level $\frac{T_{\delta}b}{a}$ and remains at that level afterwards. Similarly, the temperature level rises to $T_c$ and remains at that level afterwards. Note that the throughput and the energy consumption in Fig.~\ref{model5} are higher with respect to those in Fig.~\ref{model4}. Parallel to this observation, the monotone decrease is sharper in the power policy in Fig.~\ref{model5} compared to that in Fig.~\ref{model4}. Since the power level has to be stabilized at $\frac{T_{\delta}b}{a}$, the temperature increase cost paid for achieving certain throughput is minimized if energy consumption starts faster and drops later.

In Fig.~\ref{model6}, we set the deadline to $D=3.5$ and the energy limit to $E=17.71$. Note that this energy level is slightly less than the energy of the power policy in Fig.~\ref{model5}, which translates into a right shift of the point $t_0$. In particular, we calculate $t_0=3.2$ as the solution of (\ref{bir})-(\ref{uc}) in this case. Similar to the effect of decreasing the deadline observed in the comparison of Figs. \ref{model4} and \ref{model5}, we observe that decreasing the energy level yields a \textit{smoother} power policy. Power level drops to $\frac{T_{\delta}b}{a}$ and the temperature hits $T_c$ at a later time $t_0$ and both remain constant afterwards.

In Fig.~\ref{model7}, we consider the same system as in previous figures with two energy arrivals instead of one and with $D=5$. In particular, $E_0=6.08$ is available initially and $E_1=14.55$ arrives at time $s_1=1.5$. In this case, we calculate $t_0=3.9$. The energy causality constraint is tight and the power level makes a jump at the energy arrival instant. Note that the temperature is continuous at the energy arrival instant even though its first derivative is not. While the power level has a smooth start, a sharper decrease is observed towards the end since the harvested energy has to be fully utilized. In particular, the temperature increase before the energy arrival is kept to a minimum level so as to have a higher heat budget for the larger energy that arrives later. The temperature hits $T_c$ at $t=3.95$ after which the power and temperature both remain constant.

\begin{figure}[t]
\centerline{\includegraphics[width=0.88\linewidth]{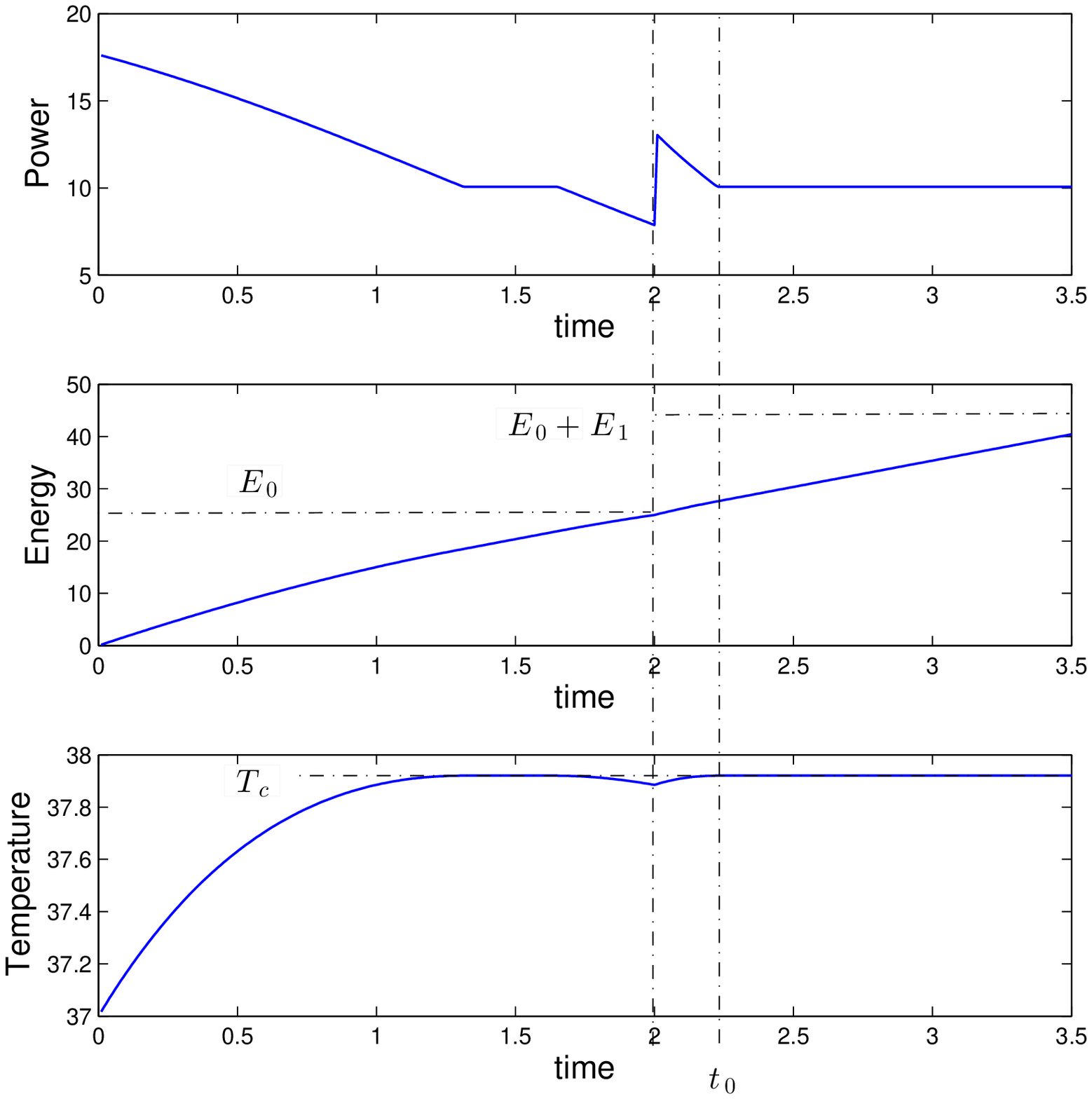}}\vspace{-0.3in}
\caption{Power, energy and temperature plots for two energy arrivals, $E_0=25$ at $t=0$ and $E_1=17$ at $t=2$ and $D=3.5$.}
\label{model8}
\vspace*{-0.1cm}
\end{figure}

Finally in Fig.~\ref{model8}, we illustrate a curious behavior in the optimal policy. For this example, we set $a=0.1$, $b=1.1$, $T_e=37$ and $T_c=37.92$. Initial energy is $E_0=25$ and energy arrives at $t=2$ with amount $E_1=17$ and the deadline is $D=3.5$. We observe that energy causality constraint is tight at $t=2$ whereas it is not tight at $t=D$ meaning that some energy is wasted in order not to cause excessive heat. The temperature generated in this throughput optimal power policy first monotonically increases, hits $T_c$ at $t=1.31$, remains there till $t=1.66$ and drops below $T_c$. We interpret the drop in the temperature in the first epoch as an effort to create temperature margin for the high energy arrival in the next epoch. We calculate $t_0=2.23$ as the time after which power level remains at $\frac{T_{\delta}b}{a}=10.12$ and the temperature remains at $T_c$. Note that under unlimited energy, temperature would hit $T_c$ at $t=0.878$. Due to the energy scarcity in the first epoch, temperature hits $T_c$ later and drops below $T_c$. A common behavior we observe in each numerical example is that temperature ultimately increases between two epochs where energy causality constraint is tight. Further research is needed to quantify the relations between the amount of temperature generated while performing optimally in terms of throughput. While monotonicity of the temperature is lost when multiple energy harvests exist, we note that monotonicity of the temperature is guaranteed in the last epoch due to Lemma~\ref{extension}.

\section{Conclusions}

We considered throughput maximization for an energy harvesting transmitter over an AWGN channel under temperature constraints. We used a linear system model for the heat dynamics and determined the throughput optimal power scheduling policy under a maximum temperature constraint by using a Lagrangian framework and the KKT optimality conditions. We determined for the single energy arrival case that the optimal power policy is monotone decreasing whereas the temperature is monotone increasing and both remain constant after the temperature hits the critical level. We then generalized the solution for the case of multiple energy arrivals. While monotonicity of the temperature is lost when multiple energy harvests exist, we observed that the temperature ultimately increases while maximizing the throughput. We also observed that the main impact of the temperature constraints is to facilitate faster energy expenditure subject to energy causality constraints. Additionally, even though using all of the available energy is optimal for throughput maximization only, with temperature constraints, energy may have to be wasted in order not to exceed the critical temperature.

\end{document}